\documentstyle[12pt]{article}

\begin{document}

\begin{titlepage}

\begin{flushright}
IJS-TP-98/22\\
NUHEP-TH-98-20\\
CPT-S6400898\\
\end{flushright}

\begin{center}
{\Large \bf 
The  CP violating asymmetry in $B^{\pm} \to M {\bar M} \pi^{\pm}$ decays\\}
\vspace{1cm}

{\large \bf B. Bajc$^{a}$, S. Fajfer$^{a,b,c}$, R. J. Oakes$^{c}$,  
T.N. Pham$^{d}$ and S. Prelov\v sek$^{a}$\\}

{\it a) J. Stefan Institute, Jamova 39, P. O. Box 3000, 1001 Ljubljana, 
Slovenia}
\vspace{.5cm}

{\it b) 
Department of Physics, University of Ljubljana, Jadranska 19, 1000 Ljubljana,
Slovenia}
\vspace{.5cm}

{\it c) Department of Physics and Astronomy, Northwestern University, 
Evanston, Il 60208, U.S.A.\\}
\vspace{.5cm}
               
{\it d) Centre de Physique Theorique, Centre National de la Recherche 
Scientifique, 
UMR 7644, Ecole Polytechnique, 91128 Palaiseau Cedex, France\\}

\end{center}

\vspace{0.25cm}

\centerline{\large \bf ABSTRACT}

\vspace{0.2cm}
We analyze  the asymmetry  in the partial widths for the 
decays 
$B^{\pm} \to M {\bar M} \pi^{\pm}$ 
($ M = \pi^+, K ^+ , \pi^0, \eta$), which results from the
interference of
the nonresonant decay amplitude with the resonant amplitude for
$B^{\pm} \to \chi_{c0} \pi^{\pm} $
followed by the decay $\chi_{c0} \to M {\bar M} $.  The CP violating phase
$\gamma$ can be extracted from the measured asymmetry.
We find that the partial width asymmetry for $B^\pm \to \pi^+ \pi^- \pi^\pm$
is about $0.33~sin \gamma$, and about $0.45~ sin \gamma$ for 
$B^\pm \to K^+ K^-\pi^\pm$, while it is somewhat smaller for 
$B^\pm \to \pi^0 \pi^0 \pi^\pm$ and $B^\pm \to \eta \eta \pi^\pm$. 
Potential sources of uncertainties in these results, primarily 
coming from poorly known input parameters, are discussed.

\end{titlepage}

The measurement  of CP asymmetries in charged B  meson decays  might provide 
the first demonstration 
of CP violation outside the K system \cite{Nir,Ali}.
Among the usual three CP-odd phases $\alpha$, $\beta$ and $\gamma$, 
the  phase $\gamma$ 
seems to be the most difficult to explore experimentally  \cite{RF}.   
One possibility to measure this CP odd phase $\gamma = arg (V_{ub}^*)$ 
has been suggested  in \cite{DEHT,EGM}. 
In this approach the asymmetry appears as
a result of the interference of the nonresonant 
$B^{\pm}\to M \bar  M \pi^{\pm}$ decay amplitude
and the resonant  $B^{\pm} \to \chi_{c0} \pi^{\pm} $ $\to M \bar  M \pi^{\pm}$ 
amplitude,
where $\chi_{c0}$ is the $0^{++}$ $c {\bar c} $ state at $3.415$ $\, \rm GeV$. 
The absorptive phase necessary to observe CP violation in the 
partial width asymmetry is provided by the $\chi_{c0}$ width.
There have also been suggestions  to determine the  
CP violating phase $\gamma$ by analyzing  the Dalitz plots  
in the decays $B^\pm \to \pi^+ \pi^- \pi^\pm$ \cite{BBGM1} and 
$B^\pm \to  \pi^- \pi^+ K^\pm$ \cite{EOS}.  
                                                    
On the experimental side the CLEO collaboration has reported 
upper limits on some of the nonresonant decays 
of the type  $B ^+ \to h^+ h^+ h^-$ \cite{CLEO}. 
CLEO found the upper limits on the
branching ratios $BR(B^+ \to \pi^+ \pi^- \pi^+) \le 4.1 \times 10^{-5}$ and
$BR(B^+ \to K^+ K^-\pi^+) \le 7.5 \times 10^{-5}$.
In  \cite{DEHT} the branching ratio for 
$B^+ \to \pi^+ \pi^- \pi^+$ was estimated to be in the range 
 $1.5 \times10^{-5}$  to $ 8.4 \times10^{-5}$.

Motivated by this theoretical expectation and the CLEO 
experimental results,
we further investigate the asymmetry in the decays 
$B^{\pm} \to M {\bar M} \pi^{\pm}$, where $ M = \pi^+, K ^+ , \pi^0, \eta$, 
improving upon the calculation of the nonresonant part of the decay 
amplitude. 
We will assume, as in \cite{DEHT},  the resonant decay amplitude in 
$B^{\pm} \to M {\bar M} \pi^{\pm}$ 
is due to the $c \bar c$ resonance 
$\chi_{c0}$  which subsequently decays 
into $M {\bar M}$, where $ M = \pi^+, K ^+ , \pi^0, \eta$.
Note that
we are interested only in the kinematical region
where the $M \bar M$ invariant mass is close to the $\chi_{c0}$ mass,
as in \cite{DEHT}. Thus $M {\bar M}$  arising from other 
resonances such as the $\rho$ 
need  not be considered.            
However, we will use  the nonresonant $B^\pm \to M  \bar M \pi^\pm$
decay amplitudes, calculated using techniques developed 
previously in our analysis of $D_{l4}$ decays \cite{BFOP}. 
In particular, we use the factorization approximation, in which 
the main contribution to the nonresonant $B^{\pm} \to M {\bar M} \pi^{\pm}$ 
amplitude comes from  the product
$< M \bar M| ({\bar u} b)_{V - A}| B^->$ $ < \pi^- | ({\bar d} u)_{V -A} |0>$
where $(\bar q_1 q_2 )_{V-A}$
denotes  $\bar q_1 \gamma_{\mu} (1- \gamma_5)q_2$.
For the calculation of the matrix element
$< M \bar M| ({\bar u} b)_{V - A}| B^->$ 
we extend  the results obtained in \cite{BFOP}, where the nonresonant
$D^+ \to K^- \pi^+ l \nu$ decay
was analyzed. In this analysis the experimental result for 
the branching ratio of 
the nonresonant $D^+ \to K^- \pi^+ l \nu$ decay was successfully reproduced
within a hybrid framework  which combines the
heavy quark effective theory (HQET)
and the chiral Lagrangian (CHPT) approach. 

The combination  of heavy quark symmetry and  chiral symmetry has been 
quite successful in the 
analysis of D meson semileptonic decays \cite{casone} - \cite{BFO}. 
The heavy quark symmetry is expected to be even  better
for the heavier B mesons \cite{caspr,wise}. However, CHPT 
could be worse in B decays due to the large
energies of light mesons in 
the final state. It is really only known  that the combination of
HQET and CHPT is valid at small
recoil momentum. In \cite{BFO} we have modified 
the hybrid model of \cite{casone} - \cite{caspr}
to describe
the semileptonic decays of D mesons to one light vector or pseudoscalar 
meson sate. Our modification is quite straightforward: we  retain 
the usual HQET Feynman rules for the
{\it vertices} near and outside the zero-recoil region, as in \cite{caspr,wise},
{\it but we include the complete
propagators instead of using the usual HQET propagator} \cite{BFO}. 
This reasonable modification  of the hybrid model enabled us to use 
it successfully  
over the entire  kinematic region of the D meson weak decays 
\cite{BFOP,BFO,BFOS}. 
The details of this approach can be found  in \cite{BFOP,BFO}.  

In the following we systematically use this model \cite{BFOP,BFO}  
to calculate the nonresonant 
$B^{\pm} \to M {\bar M} \pi^{\pm}$ decay amplitude.  We find there are 
important contributions, 
which were not taken into account  previously \cite{DEHT}.\\

The weak effective Lagrangian for the nonleptonic Cabibbo
suppressed $B$ meson decays is given by \cite{DEHT}
\begin{equation}
{\cal L}_{w} = - \frac{G_F}{{\sqrt 2}} V_{ud}^* V_{ub} (a_1^{eff} O_1 +
a_2^{eff} O_2)
\label{eq1}
\end{equation}
where $O_1 = ({\bar u} b)_{V-A} ({\bar d} u)_{V-A}$ and
$O_2 = ({\bar u} u)_{V-A} ({\bar d} b)_{V-A}$.
We take the effective coefficients $a_i^{eff}$ $(i=1,2)$ from the 
phenomenological fit 
 \cite{BHP}, which gives 
$a_1^{eff}  \simeq 1.08$ and $a_2^{eff} \simeq 0.21$. 
These values are also in agreement with other analyses \cite{N1,BSW}. 
We do not take into account the contributions arising 
from penguin operators, since these  
contributions are not expected to be important \cite{DEHT,BBGM1,DGR}.
The quark currents required in the weak Lagrangian (\ref{eq1})
can be  expressed
in terms of the meson fields, as previously described explicitly in 
\cite{BFOP,BFO}.

Using the factorization approximation, as in \cite{DEHT},  we can 
analyze all possible contributions to the $B^{\pm} \to M {\bar M} \pi^{\pm}$ 
nonresonant  amplitude.
The various kinds of  contributions are shown in Fig. 1.  
To illustrate the use of the factorization of the amplitude 
we consider the specific decay  
$B^-\to \pi^-\pi^-\pi^+$. 
It is easy to see, then,   
that the two  contributions shown in  Fig. 1a, 
which come from the operator $O_1$,   
cancel each other in the chiral limit 
$m_{\pi} \to 0$, since in this limit 
\begin{eqnarray}
< \pi^- (p_1) \pi^+ (p_2) \pi^- (p_3)| O_1 | B^- (p_B)> =  
< \pi^- (p_1) \pi^+ (p_2) \pi^- (p_3)| ({\bar d} u)_{A}| 0>& \times& \nonumber\\
 <0| ({\bar u} b)_{A} | B^- (p_B)> &+ &\nonumber\\ 
 < \pi^- (p_1) \pi^+ (p_2) \pi^- (p_3)| {\cal L}_s | \pi^-(p_B)>
\frac{-1}{m_B^2 - m_{\pi}^2} & \times &\nonumber\\
<\pi^-(p_B)|({\bar d} u )_A|0><0| ({\bar u} b)_A| B^- (p_B)>& = & 0 \nonumber\\
\label{eFig1a}
\end{eqnarray}

>From Fig. 1b there is a contribution from operator $O_2$:
\begin{eqnarray}
< \pi^- (p_1) \pi^+ (p_2) \pi^- (p_3)| O_2 | B^- (p_B)> &=& \nonumber\\ 
< \pi^- (p_1) \pi^+ (p_2) | ({\bar u} u)_V| 0>
<\pi^- (p_3)| ({\bar d} b)_V | B^- (p_B)>  &+ & \nonumber\\ 
(p_1 \leftrightarrow p_3).
 \label{eFig1b}
\end{eqnarray}
The matrix element $< \pi^- (p_1) \pi^+ (p_2) | ({\bar u} 
u)_V| 0>$ can be calculated  using  the  model developed previously 
\cite{BFO}.  
In this model the matrix element (\ref{eFig1b}) 
is  determined by the pion form factor,   
which is dominated by the $\rho$ meson pole. However, the $\rho$ meson 
contribution is not relevant in the present calculation 
of the nonresonant decay amplitude. 
Therefore, the contributions coming from 
diagrams in Fig. 1b can be neglected in our calculation of the nonresonant 
decay amplitude.

The only important contributions  
to the nonresonant decay amplitude comes 
from the diagrams in Fig. 1c and  is given by 
\begin{eqnarray}
< \pi^- (p_1) \pi^+ (p_2) \pi^- (p_3)| O_1 | B^- (p_B)>&= & \nonumber\\
< \pi^- (p_3)|( {\bar d} u)_{V-A}| 0>
< \pi^- (p_1) \pi^+ (p_2) | ({\bar u} b)_{V-A} | B^- (p_B)> & + & \nonumber\\
 (p_1 \leftrightarrow p_3). 
\label{eFig1c}
\end{eqnarray}
For the matrix element of $< \pi^- (p_1) \pi^+ (p_2) | ({\bar u} b)_{V -A} 
| B^- (p_B)>$
we use the results obtained in the analysis of the nonresonant 
$D^+ \to \pi^+ K ^- l \nu_l$ decay width \cite{BFOP}. Following this analysis 
we write the matrix element 
$< \pi^- (p_1) \pi^+ (p_2) | ({\bar u} b)_{V-A} | B^- (p_B)>$ 
in the general form 
\begin{eqnarray}
\label{wwh}
< \pi^- (p_1) \pi^+ (p_2) | {\bar u} \gamma_{\mu} (1 - \gamma_5) b | B^- (p_B)>
&\!\!\! = \!\!\!&
ir(p_B-p_2-p_1)_\mu\nonumber\\
+iw_+(p_2+p_1)_\mu+iw_-(p_2-p_1)_\mu
&\!\!\! -\!\!\!&2h\epsilon_{\mu\alpha\beta\gamma}p_B^\alpha p_2^\beta 
p_1^\gamma\;.
\end{eqnarray}
In the present  case only the nonresonant form factors $w_-^{nr}$ and 
$w_+^{nr}$ contribute. Note that the contribution proportional to $r$ is of   
order $m_{\pi}^2$ and  therefore can be 
 safely neglected. 
However, in  \cite{BBGM1}, where a different  parametrization of the form 
factors
was used,  in neglecting the  contributions of the order $m_{\pi}^2$ 
the  contributions proportional to 
$w_\pm$ were also dropped. In the notation of \cite{BBGM1}  
the product 
$<M (p_1) \bar M (p_2)|(\bar u b)_{V-A} |B(p_B)>$ 
$<\pi(p_3)|(\bar d u)_{V-A}|0>$   
is proportional to to $F_4 m_{\pi}^2$ \cite{BBMG2} and one  
can easily show that $F_4 m_{\pi}^2$ 
 $= -m_{\pi}^2 r + m_{\pi}^2/m_{\pi}^2 (w_+ + w_-) (p_B+ p_2)\cdot p_3 +$ 
$ m_{\pi}^2/m_{\pi}^2 (w_+ - w_-) (p_B + p_1)\cdot p_3 $, which  explicitly 
demonstrates that the terms proportional to 
the form factors $w_\pm$ arising from the product 
$<M\bar M|(\bar u b)_{V-A} |B>$ $<\pi|(\bar d u)_{V-A}|0>$
can not be neglected, but are important contributions to 
the nonresonant decay amplitude.  
Moreover, this contribution cannot even be  
treated as being  constant \cite{BBGM1,BBMG2}; 
it depends significantly on the variables
 $s =(p_B - p_3)^2 = (p_2+ p_1)^2$, $t =(p_B- p_1)^2 = (p_2 + p_3)^2$
 and $u = (p_B - p_2)^2 = (p_1+ p_3)^2$.

Using the preceding analysis we can write the amplitude for the 
nonresonant decay 
$B^- \to \pi^+ \pi^- \pi^-$  
\begin{eqnarray}
{\cal M}_{nr} (B^- (p_B) \to \pi^- (p_1) \pi^- (p_3) \pi^+ (p_2) )  & = &  
\frac{G_F}{{\sqrt 2}} V_{ud}^* V_{ub}\nonumber \\ 
\{a_1^{eff} [ {f_{\pi} \over 2} (m_B^2 -s -m_{\pi}^2) w_+^{nr} (s,t) 
+  {f_{\pi} \over 2} (2 t 
+ s - m_B^2 &\!\!\! - \!\!\!&3 m_{\pi}^2) w_-^{nr}(t) ] \nonumber \\
& + &  (s \leftrightarrow t)\}, 
\label{amplitude}
\end{eqnarray}
where 
\begin{eqnarray}
w_+^{nr}(s,t) & = & - \frac{g}{f_{\pi}^2} 
\frac{f_{B*} m_{B*}^{3/2} m_B^{1/2}}{t - m_{B*}^2} [ 1 - \frac{1}{2 m_{B*}^2} 
(m_B^2 -m_{\pi}^2 - t) ]\nonumber\\
& + & \frac{f_B}{ 2 f_{\pi}^2}  - \frac{{\sqrt m_B} \alpha_2}{ 2 f_{\pi}^2} 
\frac{1}{m_B^2}(2 t + s - m_B^2 - 3 m_{\pi}^2 ), 
\label{w+1}
\end{eqnarray}
and
\begin{eqnarray} 
w_-^{nr}(t) & = &  \frac{g}{f_{\pi}^2} 
\frac{f_{B*} m_{B*}^{3/2} m_B^{1/2}}{t - m_{B*}^2} 
[ 1 + \frac{1}{2 m_{B*}^2}(m_B^2 -m_{\pi}^2 - t)]
 +\frac{{\sqrt m_B} \alpha_1}{  f_{\pi}^2}.    
\label{w-1}
\end{eqnarray}
The parameters $\alpha_{1,2}$ are explicitly 
 defined in \cite{BFO}. Note that both the $\alpha_1$ and $\alpha_2$ 
terms are important in (\ref{w+1}) and (\ref{w-1}), 
which was overlooked previously \cite{DEHT}.

For the nonresonant decay amplitude 
$B^-(p_B) \to$ $  \pi^0(p_1) \pi^0(p_2) \pi^-(p_3)$ 
there are two contributions: 
one proportional to  
$a_1^{eff} <\pi^-| ({\bar d} u)_A|0>$ $< \pi^0 \pi^0|({\bar u}b)_A|B^->$  
and another one proportional to $a_2^{eff} <\pi^0| ({\bar u} u)_A|0>$ 
$ < \pi^0 \pi^-|({\bar d}b)_A|B^->$. 
The amplitude for  $B^-(p_B) \to$ $  \pi^0(p_1) \pi^0(p_2) \pi^-(p_3)$ 
is then given by 
\begin{eqnarray}
{\cal M}_{nr}( B^-(p_B) \to  \pi^0(p_1) \pi^0(p_2) \pi^-(p_3)) & = & 
\frac{G_F}{{\sqrt 2}}V_{ub} V_{ud}^* \frac{f_{\pi}}{4} \nonumber\\
\{a_1^{eff}[w_+^{nr} (s,t) (m_B^2 - m_{\pi}^2 - s) + 
w_-^{nr}(t)(2 t + s - m_B^2 & - & 3 m_{\pi}^2 ) ]\nonumber\\
+ a_2^{eff}[w_+^{nr} (u,t) (m_B^2 - m_{\pi}^2 - u) + 
w_-^{nr}(t)(2 t + u - m_B^2 & - & 3 m_{\pi}^2 )
 ]\nonumber\\
& + & (t \leftrightarrow u)\}.  
\label{aboo}
\end{eqnarray}
The form factors $w_\pm^{nr}$ appearing in the part of
amplitude  proportional to
$a_2^{eff}$ are given by (\ref{w+1}) and (\ref{w-1}), including the terms 
proportional to $\alpha_{1,2}$, 
while in the part of amplitude  proportional to
$a_1^{eff}$ these terms depending on $\alpha_{1,2}$ are absent.

A similar analysis of the nonresonant amplitude for the 
decay $B^-(p_B) \to $ $  K^+(p_1) K^-(p_2)$ $ \pi^-(p_3)$ 
gives
\begin{eqnarray}
{\cal M}_{nr} (B^- (p_B) \to K^+(p_1) K^-(p_2) \pi^- (p_3)) &= & 
\frac{G_F}{{\sqrt 2}} V_{ud}^* V_{ub} \nonumber\\
\{a_1^{eff} [ {f_{\pi} \over 2} 
(m_B^2 -  s -m_{\pi}^2) w_+^{nr} (s,t)
&\!\!\! + \!\!\!&  {f_{\pi} \over 2} (2 t + s - m_B^2\nonumber\\
 - m_{\pi}^2 & - & 2 m_K^2) w_-^{nr}(t) ] \}, 
\label{amplitudeKK}
\end{eqnarray}
where
\begin{eqnarray}
w_+^{nr}(s,t) & = & - \frac{g}{f_K^2} 
\frac{f_{B*} m_{B*}^{3/2} m_B^{1/2}}{t - m_{B*}^2} [ 1 - \frac{1}{2 m_{B*}^2} 
(m_B^2 -m_{\pi}^2 - t ) ]\nonumber\\
& + & \frac{f_B}{ 2 f_{K}^2}  - \frac{{\sqrt m_B} \alpha_2}{ 2 f_{K}^2} 
\frac{1}{m_B^2}(2 t+ s - m_B^2 - m_{\pi}^2 - 2 m_K^2), 
\label{w+1KK}
\end{eqnarray}
and 
\begin{eqnarray} 
w_-^{nr}(t) & = &  \frac{g}{f_{K}^2} 
\frac{f_{B*} m_{B*}^{3/2} m_B^{1/2}}{t - m_{B*}^2} 
[ 1 + \frac{1}{2 m_{B*}^2}(m_B^2 -m_{\pi}^2 - t)]
+  \frac{{\sqrt m_B} \alpha_1}{  f_{K}^2}.    
\label{w-1KK}
\end{eqnarray}

The analysis of the decay $B^-(p_B) \to \eta (p_1) \eta (p_2)\pi^-(p_3)$ is a 
little more complicated due to   $\eta - \eta'$ mixing.  
The nonresonant decay amplitude is 
\begin{eqnarray}
{\cal M}_{nr}( B^-(p_B) \to  \eta (p_1) \eta (p_2) \pi^-(p_3)) = 
\frac{G_F}{{\sqrt 2}} V_{ub} V_{ud}^* 
\frac{f_{K}^2}{8} [ (1 +c_\theta^2 ) \frac{1}{f_{\eta}} 
& + & s_\theta c_\theta \frac{1}{f_{\eta}'}]^2\nonumber\\
\{a_1^{eff}[w_+^{nr} (s,t) (m_B^2 - m_{\pi}^2 - s) + w_-^{nr}(t) (2 t + s 
 -  m_B^2  -   m_{\pi}^2 &  - & 2 m_{\eta}^2 )]\nonumber\\
+ a_2^{eff}[w_+^{nr} (u,t) (m_B^2 - m_{\eta}^2 - u) + w_-^{nr}(t)  
(2 t + u  -  m_B^2  -   2 m_{\pi}^2&  - & m_{\eta}^2)]\nonumber\\
 +  (t & \leftrightarrow & u)\}.
\label{abee}
\end{eqnarray}
The form factors $w_\pm^{nr}$ in (\ref{abee}) 
are given by equations (\ref{w+1KK}) and  
(\ref{w-1KK}), with $m_K \to m_{\eta}$ and without  
the terms proportional to 
$\alpha_{1,2}$. 
The $\eta - \eta'$ mixing is defined as usual with 
$\eta = \eta_8 c_{\theta} - \eta_0 s_{\theta}$ and 
$\eta' = \eta_8 s_{\theta} + \eta_0 c_{\theta}$, 
where $ c_{\theta} =cos \theta$ and  $s_{\theta}= sin \theta$ 
 and $\theta \simeq -20^0$ \cite{Aihara}. 
In the numerical calculations below we used the
values $f_{\eta} = 0.13$ $ \, \rm GeV$ and $f_{\eta '} = 0.11$ $ \, \rm GeV$ 
determined in
\cite{Aihara}. 
We will not analyze cases where the 
the $\eta'$ meson is in the final state since one expects  that 
in the nonleptonic decays of B mesons into final states 
$\eta' X$ the 
gluonic penguin contributions 
are probably  very important (see, for example \cite{AS,GR}).

The partial width for the nonresonant decay  
$B^- \to M {\bar M} \pi^- $ is 
\begin{equation} 
\Gamma_{nr} (B^- \to M {\bar M} \pi^-) = \frac{1}{(2 \pi)^3} 
\frac{1}{32 m_B^3} \int |{\cal M}_{nr}|^2~  ds~ dt. 
\label{dw}
\end{equation}
(There is an additional factor of $1/2$  
for the decays with two identical pseudoscalar mesons in the final state.)
The lower and the upper bounds on $s$ are 
$s_{min} = (m_2 + m_3)^2$, $ s_{max} = (m_B - m_1)^2$,  
while for $t$ 
they are given by
\begin{eqnarray} 
t_{min,max} (s) & = & m_1^2 + m_2^2 - \frac{1}{s} [
(s - m_B^2 + m_1^2)(s + m_2^2 - m_3^2)\nonumber\\
& \mp &\lambda^{{1\over 2}}( s, m_B^2,m_1^2)  
\lambda^{{1\over 2}}(s, m_2^2,m_3^2)], 
\label{bounds}
\end{eqnarray}
where $\lambda(a,b,c) = a^2 + b^2 + c^2 -2( ab + ac + ab)$.\\
\vspace{0.5cm}

Unfortunately there is no  experimental 
measurement of the B meson decay constant and therefore 
we will use the usual heavy quark symmetry relation \cite{caspr}  
\begin{equation}
\frac{f_B}{f_D} = \sqrt{ \frac{m_D}{m_B} }
\label{alpha}
\end{equation}
to obtain $f_B$ from the $f_D$, even though it  has not been  
experimentally measured either. 
But, there are some data on $f_{D_s}$ \cite{PDG}, 
albeit with large uncertainty, and 
taking  $f_D \simeq 200$ $ MeV$ is reasonable \cite{BFOS,GM}. Then the 
$B$ decay  constant is  $f_B \simeq 128$ $ MeV$.
We will use  the value of $|V_{ub}| = 0.0031$, the 
latest average in \cite{PDG}. 
The parameter $g$ in the form factors, determined from the 
$D^* \to D \pi$ decay,   is $g = 0.3 \pm 0.1$ \cite{PC}.
>From $D^0 \to K^- l^+ \nu$ we  found $g = 0.15 \pm 0.08$ \cite{BFOS}.
In the present calculations we  will consider the range $0.2 \leq g \leq 0.23$, 
the  overlap between these two determinations of g. 
We have previously determined the  numerical values of 
$\alpha_{1}$ and $\alpha_2$ 
 analyzing the 
$D \to V l \nu_l$ decays \cite{BFO,BFOS}, 
and must extrapolate to B mesons from these D meson values.  
To this end we apply {\it the soft scaling} of the 
axial form factors $A_1$ and $A_2$ as discussed  in \cite{caspr}.   
This scaling procedure has the virtue that it 
does not neglect the masses of the light vector mesons 
when the heavy quark symmetry is used. 
It is  completely in the same spirit as 
the basic assumption underlying our  
simple modification of the HQET propagators  
developed previously \cite{BFOP,BFO,BFOS} and  used in the present 
analysis.  
Soft scaling \cite{caspr} of the axial form factors means that     
\begin{equation}
A_1^{HV}(q_{max}^2) = {\rm const}~ \frac{{\sqrt m_H}}{m_H + m_V}
\label{a1},
\end{equation}
and 
\begin{equation}
A_2^{HV}(q_{max}^2) = {\rm const}~ \frac{m_H + m_V}{{\sqrt m_H}}, 
\label{a2}
\end{equation}
where $H$ and $V$ denote heavy pseudoscalar and light vector mesons, 
respectively.
It is easy to see that this scaling leads to the following relations: 
$\alpha_1^{DK*} = \alpha_1^{B\rho}$ and  $\alpha_2^{DK*}/m_D = 
\alpha_2^{B\rho} /m_B$.  
Among all cases found in \cite{BFO,BFOS}, we select the values 
$\alpha_1^{DK*} = -0.13$ $\, \rm GeV^{1/2}$, $\alpha_2^{DK*} = -0.13$ $\, \rm 
GeV^{1/2}$, 
since only this  choice of parameters 
gives   $3.4 \times 10^{-5}\leq$ $BR(B^- \to \pi^- \pi^+ \pi^+) \leq$
 $ 3.8 \times 10^{-5}$, 
consistent with recent data \cite{CLEO}. 
All the other combinations of $\alpha_{1,2}$, found in \cite{BFOP} give a 
$B^- \to \pi^- \pi^+ \pi^+$  branching ratio larger than 
the experimental upper limit \cite{CLEO}. 
And for this  same set of parameters, we also find
$1.4 \times 10^{-5}\leq$$BR(B^-\to K^- K^+ \pi^-) \leq$ $1.5 \times 10^{-5}$,  
which is bellow the experimental upper limit \cite{CLEO}. 
We also note the following limits for
the unmeasured branching ratios: $1.5 \times 10^{-5}\leq$
$BR(B^- \to \pi^- \pi^0 \pi^0)\leq$ $ 1.7 \times 10^{-5}$ and  
$1.0 \times 10^{-5}\leq$
$BR(B^- \to \pi^- \eta \eta)\leq$ $ 1.1\times 10^{-5}$. We note that 
the contributions to the branching ratios  arising from $\alpha_{1,2}$  
are very important in these numerical results. 

In addition to the uncertainties in our results arising from the 
uncertainties in the values of the parameters discussed above, 
there is a potentially quite large  error that could come  
from the uncertainty  in  the CKM matrix element $V_{ub}$ and 
the decay constant $f_B$. 
For example, the range of values  
$0.0018 \leq V_{ub}\leq 0.0044$ \cite{PDG} could 
change the branching ratios by as much as  a factor of 2, but 
the resulting uncertainty in 
the CP asymmetry, which we discuss next, is somewhat smaller. \\
 
\vspace{0.5cm}        

In order to obtain the  CP violating asymmetry, one also needs  to 
calculate the resonant decay amplitude $B^- \to $ $ \chi_{0c} \pi^-$
$ \to M \bar M \pi^-$. This amplitude 
can easily be determined in the narrow width approximation, 
as in \cite{DEHT}:   
\begin{eqnarray}
{\cal M}_{r}(B^- \to  \chi_{0c} \pi^- \to M \bar M \pi^-) & = &\nonumber\\
M(B^{-} \to \chi_{0c} \pi^- ) \frac{1}{ s - m_{\chi_{0c}}^2 +  i 
\Gamma_{\chi_{0c}} m_{\chi_{0c}}}
M( \chi_{0c} \to M \bar M ) + (s \leftrightarrow t).
\label{ares}
\end{eqnarray}
In our numerical calculations we will  use the estimate 
$BR(B^{\pm} \to $ $\chi_{c0}$ $ \pi^{\pm} )/$ $ BR(\chi_{c0} $ 
$\to \pi^+ \pi^-)  =$ $ 5 \times 10^{-7}$ derived in  \cite{EGM}. 
The $\chi_{c0}$ decay data \cite{PDG} 
then fix   
the decay amplitudes for 
$\chi_{c0} \to M \bar M$, ($M  = \pi^+, \pi^0, K^+, \eta$).  \\

\vspace{0.5cm}

Finally we can 
calculate the partial width  asymmetry in the 
$B^{\pm} \to M \bar M \pi^{\pm}$ decays.  
We are only interested in the kinematical region where the $M\bar M$ invariant 
mass is close to the mass of the $\chi_{c0}$ meson,  
$m_{\chi_{0c}} = 3.415~\, \rm GeV$. 
The partial decay width $\Gamma_p$ for 
$B^- \to  M \bar M  \pi^-$, which contains 
both the nonresonant and resonant 
contributions,  is obtained then by 
integration from $ s_{min} = (m_{\chi_{0c}} - 2\Gamma_{\chi_{0c} })^2$ to 
$ s_{max} = (m_{\chi_{0c}} + 2\Gamma_{\chi_{0c} })^2$, where 
 $\Gamma_{\chi_{0c} } = 0.014 \pm 0.005~\, \rm GeV$ is the width of the  $ 
\chi_{0c}$:
\begin{equation}
\Gamma_p = 
\frac{1}{(2 \pi)^3} \frac{1}{32 m_B^3} 
\int_{s_{min}}^{s_{max}} ds \int_{t_{min}(s)}^{t_{max}(s)} dt~|{\cal M}_{nr } + 
{\cal M}_{r} |^2 . 
\label{dwp}
\end{equation}
Similarly, $\Gamma_{\bar p}$, the partial decay width for 
$B^+ \to M \bar M \pi^+$, also  contains both  the nonresonant and 
resonant contributions. 
 The CP-violating asymmetry 
is defined  by 
\begin{equation}
A = |\frac{ \Gamma_p - \Gamma_{\bar p}}{\Gamma_p + \Gamma_{\bar p}}|.
\label{asym}
\end{equation}
 
For the range of values of $g$ and selected $\alpha_{1,2}$ discussed above we 
obtain the ranges 
\begin{equation}
0.33 ~ sin\gamma \leq A(B^\pm \to \pi^+ \pi^- \pi^\pm) \leq  0.34 ~sin \gamma, 
\label{ab3p}
\end{equation}
\begin{equation}
0.44 ~ sin\gamma \leq A(B^\pm \to K^+ K^- \pi^\pm)  \leq  0.45~ sin \gamma, 
\label{abKKp}
\end{equation}
\begin{equation}
0.23 ~ sin\gamma \leq A(B^\pm \to \pi^0 \pi^0 \pi^\pm)  \leq  0.24~ sin \gamma,
\label{ab00}
\end{equation}
and 
\begin{equation}
0.17 ~ sin\gamma \leq A(B^\pm \to \eta \eta \pi^\pm)  \leq  0.20~ sin \gamma.
\label{asbee}
\end{equation}
In \cite{DEHT} it was found that $A (B^\pm \to \pi^+ \pi^- \pi^\pm)$ $= 
(0.44-0.49) sin \gamma$, which differs 
from (\ref{ab3p}) due to the importance of the $\alpha_{1,2}$ terms. 

The uncertainties due to the experimental errors in
the remaining input parameters have not been included here, but we can 
roughly estimate that 
the rather large current  uncertainties in 
$V_{ub}$, $\Gamma_{\chi_{c0}}$ and $\Gamma (B^- \to \chi_{c0} \pi^-)$ could  
result in the  error in the  asymmetry being as large as even $100\%$.\\

\vspace{0.5cm}

To summarize, we have  analyzed the partial width  
asymmetry in $B^{\pm}\to M\bar M\pi^{\pm}$
decays  ($M = \pi^+$, $K^+$, $\pi^0$,  $\eta$), which signals  CP
violation, and can potentially be used to 
determine $sin \gamma$.  
The asymmetry results from the interference of the
nonresonant decay amplitude with 
the resonant decay amplitude $B^\pm \to \chi_{oc} \pi^{\pm}$ followed by
$\chi_{0c} \to M \bar M$. The asymmetry, which is rather sensitive to  
the choice of parameters, was estimated to be   $0.33~\sin\gamma$ 
for $B^\pm \to \pi^+\pi^-\pi^\pm$ and $0.45~\sin\gamma$ 
for $B^\pm \to K^-K^+ \pi^\pm $, while it is
smaller for $B^\pm \to \pi^0 \pi^0  \pi^\pm$ and  
$B^\pm \to \eta \eta  \pi^\pm$ decays. 
The estimates of these  partial width  asymmetries, 
while perhaps uncertain by as much as a factor of 2, 
do provide  useful guidance for
the experimental
searches for CP violation and a measurement of the phase  $\gamma$.

\vspace{1cm}

This work was supported in part by the Ministry of Science and Technology 
of the Republic of Slovenia (B.B., S.F., and S.P.), and by the U.S. 
Department of Energy, Division of High Energy Physics under grant 
No. DE-FG02-91-ER4086 (R.J.O.).                    
S.F. thanks the Department of Physics and Astronomy 
at Northwestern University for warm hospitality. 
\newpage 
{\bf Figure Caption}\\ 

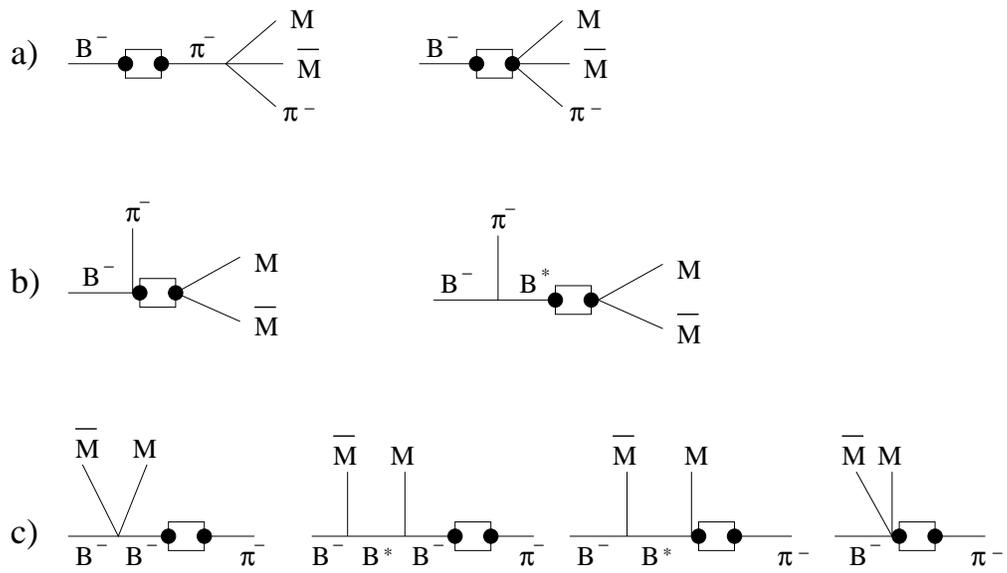
{\bf Fig. 1.} Skeleton diagrams for the  various contributions to the 
nonresonant $B^-\to M\bar M \pi^-$ amplitude. 
The square in each diagram denotes the weak transition 
due to the weak Lagrangean ${\cal L}_{w}$ (\ref{eq1}), while 
each dot denotes one of the two corresponding weak currents. \\


\newpage 
\begin{thebibliography}{10}
\bibitem{Nir} Y. Nir, talk given at 18th International Symposium on 
Lepton Photon Interactions, Hamburg, Germany, July 28 - August 1 1997, 
hep-ph/9709301.
\bibitem{Ali} A. Ali, G. Kramer and Dai-Dian L\" u, 
hep-ph/9805403.
\bibitem{RF} R. Fleisher, Talk given at International 
Euroconference QCD 98, Montpellier, France, 2-8 July 1998, hep-ph/9808238. 
\bibitem{DEHT} N. G. Deshpande, G. Eilam, X. G. He and J. 
Trampeti\' c, 
Phys. Rev. D {\bf  52} (1995), 5354.
\bibitem{EGM} G. Eilam, M. Gronau and R. R. Mendel, 
Phys. Rev. Lett. {\bf 74} (1995), 4984.
\bibitem{BBGM1} I. Bediega, R. E. Blanco, C. G\" obel, and R. Mendez-Gelain,   
et al, hep-ph/9804222.    
\bibitem{EOS} R. Enomoto, Y. Okada, and Y. Shimizu, 
 Phys. Lett. B {\bf 433} (1998), 109.
\bibitem{CLEO} T. Bergfeld et al., Phys. Rev. Lett. {\bf 77} (1996), 4503.
\bibitem{PDG} Review of Particle Physics, Eur. Phys. Jour. C {\bf 3} 
(1998), 1.                                                            
\bibitem{BFOP}B. Bajc, S. Fajfer, R. J. Oakes and T. N. Pham, 
Phys. Rev. D. {\bf 58} (1998), 054009.  
\bibitem{casone} R. Casalbuoni, A. Deandrea, N. Di Bartolomeo, R. Gatto, 
F. Feruglio and G. Nardulli, Phys. Lett. B {\bf 292} (1992), 371.
\bibitem{castwo} R. Casalbuoni, A. Deandrea, 
N. Di Bartolomeo, R. Gatto, F. Feruglio and 
G. Nardulli, Phys. Lett. B {\bf 299} (1993),  139.
\bibitem{casthree} A. Deandra, N. Di Bartolomeo, R. Gatto 
and G. Nardulli, Phys. Lett. B {\bf 318} (1993), 549.
\bibitem{caspr} R. Casalbuoni, A. Deandrea, N. Di Bartolomeo, R. Gatto, 
F. Feruglio and G. Nardulli, Phys. Rep. {\bf 281} (1997), 145.
\bibitem{wise} M. B. Wise, Phys. Rev. D {\bf 45} (1991) (1991), Phys. Rev.
D {\bf 44} (1991), 3567. 
\bibitem{BFO} B. Bajc, S. Fajfer and R. J. Oakes, 
Phys. Rev. D {\bf 53} (1996), 4957. 
\bibitem{BFOS} B. Bajc, S. Fajfer, R. J. Oakes and S. Prelov\v sek, 
Phys. Rev D  {\bf 56} (1997), 7027.
\bibitem{BHP} T. E. Browder, 
K. Honsheid, and D. Pedrini, An. Rev. Nucl. Part. Sci. {\bf 46} (1997), 
395.
\bibitem{N1} M. Neubert, Inv. talk presented at the EPS meeting, 
Jerusalem, Israel, 19-25 Aug. 1997, hep-ph/9801269.
\bibitem{BSW} M. Bauer, B. Stech, and M. Wirbel, Z. Phys. C {\bf 34} (1987), 
 103.
\bibitem{DGR} A. S. Dighe, M. Gronau and J. L. Rosner, 
hep-ph/9709223.
\bibitem{BBMG2} I. Bediega, R. E. Blanco, C G\" obel and R. Mendez-Gelain, 
Phys. Rev. Lett. {\bf 78} (1997) 22.
\bibitem{ourpaper} B. Bajc, S. Fajfer and R.J. Oakes, 
Phys. Rev. D {\bf 51} (1995),  2230.
\bibitem{Aihara} H. Aihara et al., Phys. Rev. Lett. {\bf 64} (1990),  172. 
\bibitem{AS} D. Atwood and A. Soni, Phys. Lett. B {\bf 405} (1997),  150. 
\bibitem{GR}M. Gronau and J. Rosner, Phys. Rev. D {\bf 53} (1997), 2516.
\bibitem{GM} G. Martinelli, Nucl. Inst. Meth. Phys. Res. A {\bf 384} 
(1996), 241.  
\bibitem{PC} P. Colangelo, talk given at {\it Hyperons, Charm and Beauty 
Hadrons}, 30 June - 3 July, 1998, Genoa, Italy. 

\end{thebibliography}
\end{document}